\documentclass[10pt,a4paper]{article}
\usepackage[utf8]{inputenc}
\usepackage[english]{babel}
\usepackage{amsmath}
\usepackage[pdftex]{graphicx}
\usepackage{epstopdf}
\usepackage{caption}
\usepackage{multicol}
\usepackage{subcaption}
\usepackage{times}
\usepackage{hyperref}
\usepackage{float}
\usepackage{indentfirst}
\usepackage{geometry}
\usepackage{array}        

\begin{document}

\begin{center}
\fontsize{14}{16}\selectfont
\noindent \textbf{INCREASING THE OPENING SPEED OF THE PLASMA OPENING SWITCH ON AN DIRECT ACTION ACCELERATOR WITH AN INDUCTIVE ENERGY STORAGE DEVICE}
\end{center}

\begin{center}
\fontsize{12}{14}\selectfont
\noindent \textbf{D.V. Vinnikov, O.M. Ozerov, V.V. Katrechko, V.I. Tkachov, O.V. Manuilenko, I.N. Onishchenko}

\end{center}
\begin{center}
\fontsize{10}{12}\selectfont
\noindent \textbf{\emph{National Science Center "Kharkiv Institute of Physics and Technology", Kharkiv, Ukraine}}

\textbf{\emph{E-mail: onish@kipt.kharkov.ua}}
\end{center}

\textbf{Problem.} To increase the voltage multiplication factor in a small-sized direct-acting electron accelerator DIN-2K with an inductive energy storage and a plasma opening switch, it is necessary to ensure an increase in the rate of change of the current and its amplitude during the POS opening for the purpose of obtaining an explosive electron emission with the formation of an electron beam and a virtual cathode. However, the opening process depends on many electrical parameters of the plant, and it re-quires determining their joint and individual effect on its dynamics. The purpose of this research is to study and determine the effects of electrical parameters, including those of the discharge voltages of the entire electrical circuit of the accelerator on the dynamics of the plasma opening switch, in particular opening speed, current amplitudes, and opening time, as well as to give recommendations on optimizing the operation of accelerators of this type and highlight possible ways to increase the voltage multiplication factor. Methodology. A method for determining the induced voltage according to experimental current oscillograms has been proposed. The methodology was verified by the coincidence of its data with the results of measurements by a capacitive voltage divider with the data spread of less than 20\%.
Scientific novelty. The rates of change in the current at the plasma switch opening stage were determined depending on the main electrical parameters of the DIN-2K accelerator. The diagnostics of the voltage induced during the POS opening were provided using no capacitive voltage divider in the internal volume of the accelerator, which enabled the removal of some diagnostic tools from the working volume of the chamber. Practical significance. A verified method for determining the rate of change in current during the POS opening using experimental oscillograms has been proposed, which enabled the implementation of the unhindered flow of the entire sequence of physical processes. Practical recommendations are proposed for choosing the values of electrical parameters that ensure an increase in the rate of change of current during the POS opening. Bibliography. 18, Fig. 9.


\begin{flushleft}
\fontsize{10}{12}\selectfont
\noindent \textbf{USC:}  621.384.64; 621.3.038.624
 \end{flushleft}

\columnseprule=0pt\columnsep=24pt
\begin{multicols}{2}

Accelerator technology is now an integral part of many science-intensive, high-tech processes that require high energy concentration, in particular 100 keV and above in an electron beam, and 10 MW and above power as electromagnetic radiation in a short period of time of 100 ns and less [1-11]. The areas of use of accelerators are expanding from purely physical experiments to applications in X-ray technology, radio engineering, medicine, the food industry and in the military. In this regard, much attention is paid to the development and improvement of compact, mobile installations that have high efficiency, operate in the frequency regime and have a large frequency resource.
Typically, direct acting electron accelerators operating to create electron beams and electromagnetic radiation generated by a virtual cathode are rather bulky structures, in which the main structural space is occupied by a high-voltage storage system of a capacitive type. The accelerators based on Marx generators and Tesla transformers with oil or water forming lines [5, 12, 13] and using magnetic solenoids [14] constitute most of the research fund of available laboratories and research institutions [1-11]. Such types of accelerators are bulky and heavy and require additional synchronization of capacitive sections, the use of many dischargers and the arrangement of high-voltage units in oil. Accelerators based on inductive storage can serve as an alternative to power supply circuits with capacitive energy storage, however, one of the main reasons limiting the use of circuits with inductive storage is the release of energy in a short period of time and the fact that this is a more complex engineering task than creating a circuit breaker [15-21].
National Science Center “Kharkiv Institute of Physics and Technology” is involved in the development and improvement of the small-sized direct-acting electron accelerators (DAEAs) with a typical volume of about 104 cm3, the operation of which is based on the principle of inductive energy storage (IES) and its release when the plasma opening switch (POS) is opened. [10, 11]. To do this research, experiments were carried out using the DIN-2K accelerator, which is a high-current high-voltage unit. The main advantages of this type of accelerator are its significantly smaller mass and dimensions, structural simplicity, mobility and compactness, as well as the unavailability of the usually bulky Marx generator as a power supply system, and there is also no need in a magnetic system required for focusing the electron beam. These advantages are possible due to the use of the principle of the fast opening of the current cord in the POS plasma to obtain a high voltage on the accelerator diode, in-stead of the use of the pulsed high-voltage generators. DAEAs with IES and POS are characterized by a wide range of physical phenomena, namely the formation of a plasma bridge and its rupture, voltage induction, the occurrence of explosive electron emission, the formation and propagation of a high-current electron beam (HEB) and plasma formed by the beam, the formation of a virtual cathode (VC), the generation of X-ray and electromagnetic radiation (EMR) and, accordingly, a wide range of vital practical applications and theoretical research. [1-3, 22-25]. Innovative vircator systems are also developed [3, 27,  28].
A wide range of applications varied from pure re-search to use in technologies of critical infrastructure industries impart this investigation additional significance.
The POS opening process is decisive, because particularly the release rate and the amount of energy stored in the inductive accumulator determine the im-plementation of the entire chain of subsequent physical processes, ensuring the maximum multiplication factors of the induced voltage.
The shorter the opening time and the higher the rate of increase in the resistance of the POS, the higher the efficiency of energy transfer to the load [29-36]. Despite all these significant advantages of accelerators with IES and POS, the opening stage is still a weak point in the accelerators of this type. Ensuring an in-crease in the POS opening rate and the current change amplitude is an important scientific and technical task.
Increasing the efficiency factor and optimizing the operation of the DAEA with IES and POS can be ensured by adjusting the parameters of the discharge circuits of the bank of plasma guns (PG) and the main discharge circuit that feeds the coaxial IES. The optimal parameters are those that provide the maximum amplitude and the shortest time values of the opening current of the POS, which results in the induction of an increased voltage applied to the diode.
The purpose of the research is to study and determine the effect of electrical parameters, including the discharge voltages of the circuits of the complete electrical circuit of the accelerator, on the dynamics of the plasma opening switch, in particular opening speed, current amplitude, opening time, and to provide recommendations for optimizing the operation of accelerators of this type and possible ways to increase the voltage multiplication factor.

\begin{center}
\fontsize{11}{14}\selectfont
\noindent \textbf{Principle of operation of DAEA with INE and POS.}
\end{center}

From the point of view of electrical engineering, the accelerator consists of two discharge circuits isolated from each other, see Fig, including circuit of plasma guns that generate plasma to form a current cord in it that functions as the POS. The circuit is powered by PCG-2. The main discharge circuit is powered by its pulse current generator PCG-1. The current of this circuit flows in two ways. The first $I_{1}$ flows through the plasma formed by the guns, the second $I_{2}$ is formed when POS is opened and flows through the load in the form of a vacuum diode (VD), consisting of a cathode and an anode mesh.

In terms of the operating time, the first discharge circuit of the guns ensures the creation of plasma in the evacuated volume. Six Bostik coaxial-type plasma guns with the discharge of a surface type on the dielectric serve as the load. Each gun is powered by a capacitive accumulator IK-100/0.4. Each storage circuit of the plasma guns is charged through the switching inductive chokes.

\begin{figure}[H]
  \centering
  \includegraphics[width=0.45\textwidth]{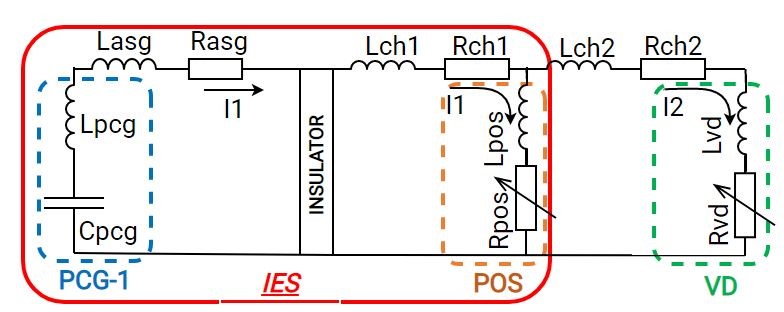}\\
  \caption{\emph{Electrical equivalent circuit of the accelerator. Cpcg and Lpcg are inductance and capacitance of the pulsed current genera-tor PCG-1; Lasg are Rasg are the relevant parameters of the air-controlled discharger; Lch1, Rch1 are inductance and re-sistance of the coaxial chamber taking into account the POS; Lch2, Rch2 are inductance and resistance of the coaxial chamber after the POS; Lpos, Rpos are the relevant parameters of the POS; Lvd, Rvd are the relevant parameters of the vacu-um diode.}}\label{fig:image_label}
\end{figure}

\noindent Electrical breakdown on the dielectric surface and its simultaneity occur due to a common air-controlled discharger and the same length of the high-voltage power cables.
And in terms of operating time, the second electrical circuit is the main one, because the voltage sup-plied from its pulse current generator to the coaxial vacuum diode of the accelerator ensures the flow of current through the plasma created by the plasma guns, the formation of a current cord, which subsequently moves through the plasma and is opened as a result of a series of physical processes [22].
When the current flowing through the plasma is opened, the energy stored in the inductance of the co-axial system is released in the load, the resistance is abruptly increased approximately by two orders of magnitude, and this, accordingly, results in a voltage jump and subsequent explosive emission of electrons from the end of the cathode and the closing of the VD by it. Then, the current flows along the coaxial circuit, closing on the beam-plasma load formed in the anode-cathode gap of the diode.
Thus, the equivalent circuit of the accelerator shown in Fig. 1 consists of two loops that are electrically interconnected due to dynamic physical process-es in the diode.
On completion of the energy storage in PCG-2, see Fig. 2 and the operation of the discharger P2, the plasma 12 created by the surface discharge of the guns 6 is injected into the accelerator volume. The energy stored in the PCG-1, with a set delay time, is fed to the load in the form of a coaxial system of electrodes 3,4 through an air-controlled switching discharger. The load consists of the POS and VD, which are involved in series. The current I1 is limited by the location of the plasma injected by the plasma guns. The inductance that affects the amount of stored energy in the INE 10 is formed from the inductance of the PCG Lpcg, which includes the inductances of the coaxial high-voltage connecting cables, the capacitor, the air-controlled discharger Lasg and that part of the coaxial electrode system Lch1 that is located to the left of the formed current cord in the POS 7 and the inductance of the POS Lpos.itself. This circuit operates until the POS is triggered.

\begin{figure}[H]
  \centering
  \includegraphics[width=0.45\textwidth]{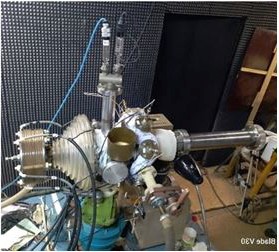}\\
\end{figure}
\begin{figure}[H]
  \centering
  \includegraphics[width=0.45\textwidth]{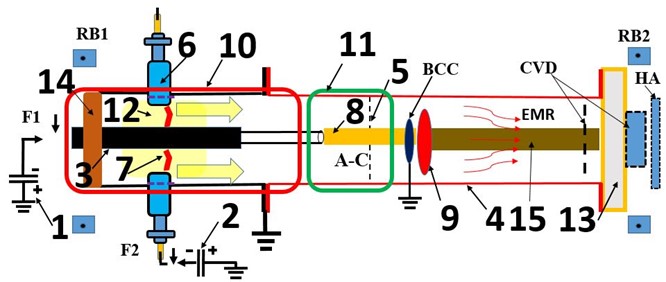}\\
  \caption{\emph{General view of the small-sized direct-acting electronic accelerator DIN-2K and its schematic representation. 1 – PCG-1 of the main discharge circuit; 2 –PCG-2 of the bank of plasma guns; 3 – tubular cathode; 4 – grounded housing of the accelerator that acts as the anode, 5 – anode mesh; 6 – plasma guns; 7 – conventional image of the current cord; 8 – high-current electron beam; 9 – virtual cathode (VC); 10 – IES inductive energy storage unit; 11 – vacuum diode; 12 – plas-ma formed by plasma guns; 13 – plexiglass end flange; 14 – high-voltage insulator; 15 – beam after passing through the VC; RB1, RB2 - Rogowski belts; F1, F2 – air-controlled dis-chargers; EMR – electromagnetic radiation; A-C – anode-cathode gap of a vacuum diode; BCC – beam current collec-tor; CVD – capacitive voltage divider; HA – EMR recording horn antenna.}}\label{fig:image_label}
\end{figure}

After the POS is triggered, the plasma conductivity is abruptly disturbed and the current $I_{1}$ begins to flow through the vacuum diode and is designated as $I_{2}$, where the VD is formed by the tubular cathode 3 and the anode mesh 5 (A-C), where an explosive emission of electrons occurs from the end of the cathode as a result of a sudden increase in voltage, and a high-current electron beam 8 is formed. Due to the volume charge of the HEB behind the anode mesh of the vacuum diode, a virtual cathode (VC) 9 may be formed, which is a source of microwave radiation, which is distributed in the waveguide 4 and goes outside. This fast process lasts on average for 100 ns. Thereafter, the remaining energy is released through the plasma remaining after the POS opening and the plasma formed by the beam 15, see Fig. 3 (oscillogram).
Fig. 3. shows a typical oscillogram of the current obtained from RB1, flowing in the inductive storage through the coaxial system of accelerator electrodes and through the POS. The energy storage time is limited by the time of the POS existence before its open-ing. In the case shown in the oscillogram, it is equal to 760 ns. The greater the current amplitude before the moment of opening, the higher the energy that can be released during the opening itself. The stage of conduction disturbance during which the current is rapidly decreased, and the voltage is abruptly increased lasts up to 70 ns. During the opening, the current flows through the circuit in the vacuum diode. A sufficient value of the induced voltage Uind allows for an explosive emission, which initiates the formation of an electron beam Ibeam, with the current of 8 kA. The recalculated induced voltage is 190 kV, which is indicative of the greatest possible energy obtained by the electrons of the HEB. After the beam is defocused and the plasma conductivity is partially restored, secondary discharge processes occur. The accumulator, on condition that there is some energy left in it, continues to discharge through the plasma that fills the coaxial part of the electrodes and the vacuum diode.

\begin{figure}[H]
  \centering
  \includegraphics[width=0.45\textwidth]{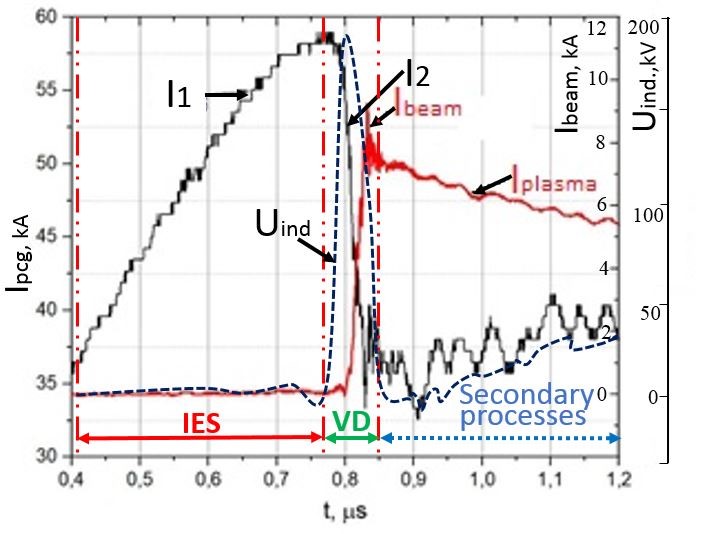}\\
  \caption{\emph{Oscillogram of the current in the inductive energy storage, the POS opening current, the beam current formed in the vacuum diode and the induced voltage estimated from the current drop rate.}}\label{fig:image_label}
\end{figure}

\begin{center}
\fontsize{11}{14}\selectfont
\noindent \textbf{Experimental equipment and methods.}
\end{center}

The charging voltage was determined by kilovoltmeters of a C-196 type. Data recording was carried out using Siglent SDS 2204X, 200 MHz and OWON XDS3104E, 100 MHz oscilloscopes. The pulsed triggering unit for 18 kV. The G5-15 pulse generator-based delay unit with a delay time of 100 ns-500 $\mu$s with a time step of 100 ns.
Rogowski belts with a bandwidth ranging from 1.6 MHz to 994.7 MHz were used for measuring the beam current and the current of the main discharge circuit. The induced voltage was determined by the calculation method from the experimental current oscillogram using the rates of change in current and inductance. The induced voltage was derived from the formula:

\begin{equation}
    U_{ind} = -L {\frac {dI} {dt}}
\end{equation}

Where $U_{ind}$ is the voltage, L is the inductance, ${\frac {dI} {dt}}$ is the current derivative in terms of time, which is determined from the oscillogram. L is determined from the experimental current oscillograms [11] by the Thomson formula:

\begin{equation}
    L =  {\frac {T^2} {4{\pi}^{2} C}}
\end{equation}

where L is the inductance, T is the oscillation period; C is the capacity of the PCG-1 storage battery.
The certified capacitive voltage divider PVM-1 manufactured by the North Star High Voltage Company. The current beam collector in the form of a Faraday cylinder. Capacitive voltage divider with a high-voltage arm capacitance of 4.2 pF, and a low-voltage arm capacitance of 21 nF. The conversion coefficient was 5 kv/V. Trapezoidal horn antenna, with a cutoff frequency of 2.1 GHz, see Fig. 2. The frequency of microwave radiation was controlled by the size of the anode-cathode gap of the vacuum diode and the discharge voltage of the PCG-1.
Fig. 4. shows the dependences of the induced voltage on the discharge voltage of the PCG-1. Curve number 1 was constructed according to the data obtained from the capacitive voltage divider (CVD) at three points for each value of the discharge voltage, see Fig. 2. The CVD was placed inside the accelerator chamber, and it was calibrated using the certified voltage divider PVM-1, and the data error was within 7\%.
The measurements were taken in the range of discharge voltages of the main discharge circuit of Upcg varying from 25 kV to 40 kV at the same voltage of 16 kV on the guns. It should be noted that the measurements of the induced voltage were significantly affected by parasitic capacitances relative to the ground, which could be an additional source of the total error. Curve number 2 corresponds to the data obtained by recalculating the voltage using the formulas 1, 2 at three points for each value of the discharge voltage of the main circuit in the range of 15 kV to 50 kV, with the discharge voltage on the plasma guns Upg = 16 kV and other equal conditions. Polynomial curves were constructed for each case based on the obtained data. The error between the two methods of voltage measurement was within 20\%
 in the voltage comparison range.

\begin{figure}[H]
  \centering
  \includegraphics[width=0.45\textwidth]{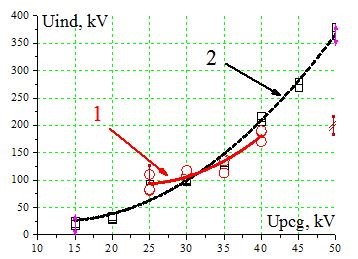}\\
  \caption{\emph{Dependence of the induced voltage on the discharge voltage of the main discharge circuit. 1 –the capacitive voltage divider data; 2 – the current oscillogram data.}}\label{fig:image_label}
\end{figure}

Each series of experiments consisted of the data package where the final data of one experimental point were selected from a sample of 5 to 10 pulses for each of the parameters studied. The most characteristic data were selected from the resulting sample. The statistical data spread for one sample does not exceed 15\%.
 The experiments were carried out to investigate the effect of the rate of change in current during the POS operation on the main electrical parameters of the accelerator. The discharge voltage of the main circuit varied from 15 to 45 kV, in increments of 5 kV. The discharge voltage of the plasma gun circuit varied from 5 to 20 kV, in increments of 5 kV. The discharge voltage of the main circuit the plasma guns circuit was fixed and was to 35 kV and 16 kV, respectively, while changing other parameters. Vacuum conditions are not worse than $3\cdot10^{-3}$ Pa.

To determine the effect of the amount of plasma injected by the guns into the accelerator volume, the number of guns varied from 1 to 6. To determine the effect of inductance on POS dynamics, we varied the number of high-voltage power cables connecting the capacitive storage of the PCG of the main circuit with the discharger P1.

\begin{center}
\fontsize{11}{14}\selectfont
\noindent \textbf{Results of the experimental studies of the POS opening.}
\end{center}

The dependences of the change in current on time when changing the main electrical parameters of the accelerator have been plotted. Fig. 5. shows the dynamics of the current flowing through the main discharge circuit of the accelerator before, during and after the POS opening depending on the discharge voltage of the main circuit powered by the PCG-1. Fig. 5. shows the current sections on the oscillograms where the opening of the POS occurs. Below, the same section is shown on an enlarged scale, with current curves, starting from the point of the opening origination. Similar sections were constructed and analyzed when studying changes in other electrical parameters of the accelerator. Fig. 5 shows that the process of energy storage in the IES occurs faster with an increase in the discharge voltage. The energy accumulation time at 45 kV is 1.5 times less than at 15 kV. At the same time, the amplitude values of the current in the IES are 2.2 times higher, at the same values of the discharge voltage. The current drop differs by 2.5 times, respectively. It is decreased 4.5 times, which is the most important for inducing the increased voltage and it also defines its multiplication factor.

Fig. 5. Dependence of the current dynamics of the main discharge circuit for the values of the discharge voltage of the PCG-1 varying in the range of 15 kV to 45 kV.
The amplitude of the POS opening current is equal to $\bigtriangleup$I and it is calculated as the difference between the maximum and minimum current values at the stage of the POS opening. The time $\bigtriangleup$t, taken for the POS open-ing is determined by the voltage from the current curve [10].

 The point of the beginning of its jump-like growth and the lowest point on its decline determine the duration of this physical process. The induced voltage was derived from formula (1), where the voltage multiplication factor is equal to the ratio of the amplitude value of the induced voltage to the discharge voltage of the main circuit.
Based on the analysis data of the POS opening section, see Fig. 5., the dependence of the current change rate on the discharge voltage of the PCG of the main circuit was plotted, as shown in Fig. 6. The figure shows that there is a gradual increase in speed for two sections with different angles of inclination, shown by an auxiliary dotted line. The breakdown occurs at the boundary of 30 to 35 kV. After 35 kV, physical processes are accelerated 5 times.

\begin{figure}[H]
  \centering
  \includegraphics[width=0.45\textwidth]{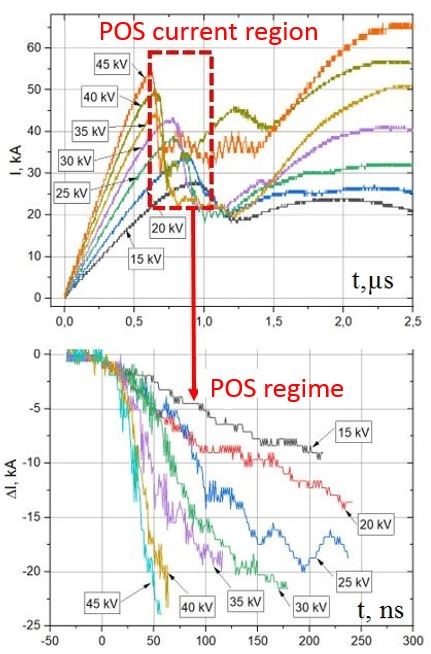}\\
  \caption{\emph{Dependence of a change in current on the discharge voltage of the PCG-1 during the POS opening.}}\label{fig:image_label}
\end{figure}

\begin{figure}[H]
  \centering
  \includegraphics[width=0.45\textwidth]{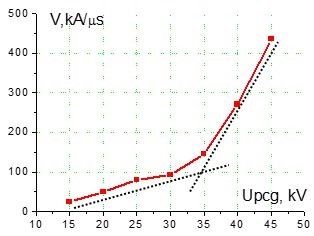}\\
  \caption{\emph{Dependence of the rate of change in current during the POS opening on the discharge voltage of the main discharge circuit.}}\label{fig:image_label}
\end{figure}

\begin{figure}[H]
  \centering
  \includegraphics[width=0.45\textwidth]{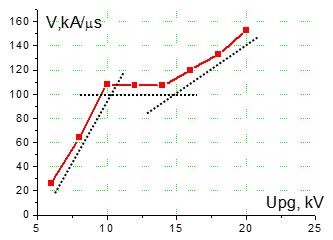}\\
  \caption{\emph{Dependence of the rate of change in current during the POS opening on the discharge voltage of plasma guns.}}\label{fig:image_label}
\end{figure}

The voltage applied to the plasma guns affects the density of the plasma and the degree of its ionization, see Fig. 7. Three characteristic areas are observed. Two with a linear increase in speed and the central area in the range of 10 to 14 kV, which is characterized by a horizontal plateau. As the voltage on the guns is increased, they form a plasma of a higher density. Depending on the density of the plasma, there is such a limiting current that the plasma can still allow to pass [22]. With an increase in voltage, the rate of change of the current speed slows down, which is visible from the slope angles of the sections of the experimental curve, i.e., the opening of the POS in a plasma with a higher density gradually becomes more difficult.

\begin{figure}[H]
  \centering
  \includegraphics[width=0.45\textwidth]{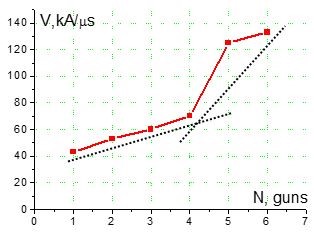}\\
  \caption{\emph{Dependence of the rate of change in current on the number of plasma guns during the POS opening.}}\label{fig:image_label}
\end{figure}

The number of plasma guns determines the amount of plasma injected into the volume. Fig. 8 shows that a break is observed at the number of guns equal to four. A smaller number of guns fails to ensure uniform filling of the volume with plasma. Five and/or six guns result in a jump-like 2.2-fold increase in the rate of change in current. A larger amount of plasma can pass a larger current resulting in larger values of dI.

\begin{figure}[H]
  \centering
  \includegraphics[width=0.45\textwidth]{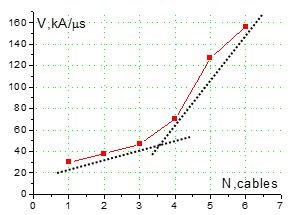}\\
  \caption{\emph{The dependence of the rate of change in current on the number of high voltage connecting cables during the POS opening.}}\label{fig:image_label}
\end{figure}

The number of high-voltage cables connecting the discharger P1 with the PCG storage determines the effect of the inductive component on the POS opening rate. The cables are arranged in parallel and, accordingly, their addition reduces the total induction of the accelerator system. It is obvious that an increase in the number of connecting cables and a decrease in their length results in a decrease in the inductance of the system and, accordingly, in an increase in the rate of operation of the circuit breaker, see Fig.9. An increase in the number of cables from 1 to 6 provides an increase in the rate of change in the current of 5.3 times. The total inductance with one cable is 1005 nHn, and with six cables it is 390 nHn. The dependence tends to the linear growth, consisting of two sections with different slope angles

\begin{center}
\fontsize{11}{14}\selectfont
\noindent \textbf{Conclusions.}
\end{center}

To ensure fast ($\sim$100 ns) POS opening for small-sized direct-acting electron accelerators with inductive energy storage and a plasma opening switch with parameters close to DIN-2K, the following conditions should be met for choosing electrical parameters:

\noindent 1. Ensuring an increase in the discharge voltage of the main electrical circuit. For DIN-2K, this voltage should be higher than 30 kV. Increasing the voltage results in a shortened energy storage time in the inductive storage, an increase in dI values and a decrease in dt values during the POS opening. The opening speed, ${\frac {dI} {dt}}$ is increased.

\noindent 2. Ensuring an increase in the number of plasma guns, providing their symmetrical arrangement relative to the cathode. Compliance with these requirements allows a greater current to be passed through the plasma and thus obtain larger dI values. For DIN-2K, adding a gun results in an increase of the main circuit current by 2.5 to 3 kA.

\noindent 3. Choosing the value of the discharge voltage of the electric circuit of plasma guns at such a level that ensures reliable POS opening by the PCG current of the main circuit. For DIN-2K, the plasma density created by plasma guns with a discharge voltage of more than 18 kV cannot provide the POS opening at the main circuit voltage less than 25 kV.

\noindent 4. Creating low-inductive discharge circuits, which ensures an increase in both the maximum values of the discharge current and  ${\frac {dI} {dt}}$.

\noindent 5. Using the DIN-2K accelerator as an example, it was possible to ensure an increase in the output parameters of the maximum discharge current, the rate of opening and the induced voltage and thus ensure reliable conditions for stable operation of the accelerator to create high-current electron beams.

\noindent 6. The units of this class may be in demand for use in many industrial sectors and in the field of cutting-edge research, for example, for the generation of electromagnetic radiation.

\textbf{Conflict of interest.} The authors declare that there is no conflict of interest.

\begin{center}
\fontsize{11}{14}\selectfont
\noindent \textbf{REFERENCES}
\end{center}

\noindent [1]	Benford, J.A. Swegle, E. Schamiloglu. High power microwaves. CRC Press, 2015, 447 p.

\noindent [2]	James Benford, Edl Schamiloglu, Jacob Coty Stephens, John A. Swegle, Peng Zhang. High Power Microwaves. Engineering  Technology, Physical Sciences. pp. 510, 2024. DOI: 10.1201/9781003287704.

\noindent [3]	S. Mumtaz, H.S. Uhm, E.H. Choi. Progress in vircators towards high efficiency: Present state and future prospects // Physics Reports. 2024, v. 1069, p. 1-46. DOI: 10.1016/j.physrep.2024.03.003.

\noindent [4]	Ilario Boscolo, M. Leo,Armando Luches, L. Provenzano. Elec-tron beam production by a Tesla transformer accelerator. Review of Scientific Instruments 48(7):747 – 751. 1977. DOI: 10.1063/1.1135155.

\noindent [5]	Tsukasa Nakamura,Motohiro Teramae, Fumiya Niwa, Hiroaki Ito. Output Evaluation of Microwave Pulse Emitted from Axially-Extracted Vircator with Resonance Cavity / Hasegawa, Jun (ed.). Tokyo Institute of Technology, Tokyo (Japan), Jan 2018, p. 55-60.

\noindent [6]	H. Zhang, T. Shu, S. Liu, Z. Zhang, L. Song, H. Zhan. A com-pact 4 GW pulse generator based on pulse forming network-Marx for high-power microwave generation // Rev. Sci. Instrum. 2021, v. 92, p. 064707; DOI: 10.1063/5.0040111.

\noindent [7]	Sohail Mumtaz, Eun-Ha Choi. The Characteristics of the Second and Third Virtual Cathodes in an Axial Vircator for the Generation of High-Power Microwaves. Electronics 2022, 11(23), 973; https://doi.org/10.3390/electronics11233973.

\noindent [8]	Se-Hoon Kim, Chang-Jin Lee, Wan-Il Kim, Kwang-Cheol Ko, "Investigation of an Axial Virtual Cathode Oscillator with an Open-Ended Coaxial Cathode ", Journal of Electromagnetic Engineering and Science, vol.22, no.3, pp.265, 2022. DOI: 10.26866/jees.2022.3.r.86.

\noindent [9]	Sohail Mumtaz, Linh Nhat Nguyen, Hansup Uhm, Pradeep Lamichhane, Suck Woo Lee, Eun Ha Choi. A novel approach to form second virtual cathode by installing a floating zone plate inside the drift tube. Results in Physics 17 (2020) 103052. pp. 1-8. https://doi.org/10.1016/j.rinp.2020.103052.

\noindent [10] D.V. Vinnikov, O.M. Ozerov, V.V. Katrechko, V.I. Tkachov, S.V. Marchenko, B.O. Brovkin, V.B. Yuferov, O.V. Manuilenko, I.N. Onishchenko. Conditions ensuring maximum voltage multiplication factors for the din-2k accelerator. Problems of Atomic Science and Technology. 2025. N1(155) Series: Plasma Physics (31), p. 69-74. DOI: 10.46813/2025-155-069.

\noindent [11]	D.V. Vinnikov, O.M. Ozerov, V.V. Katrechko, V.I. Tkachov, S.V. Marchenko, V.B. Yuferov, O.V. Manuilenko, I.N. Onishchenko. Experimental study of plasma opening switch in electron accelerator with inductive energy storage. Problems of Atomic Science and Technology. 2025. N4(158). DOI: 10.46813/2025-158-003.

\noindent [12]	Sunil Kanchi1, Rohit Shukla, Archana Sharma. Simulation and experimental results of plasma opening switch operation with inductive and electron beam diode loads in different plasma regimes. Physica Scripta, Volume 99, Number 2. 2024. 025614. DOI 10.1088/1402-4896/ad1d9d.

\noindent [13] R.Shukla. S.K.Sharma, P.Banerjee, P.Deb, T. Prabaharan, R. Das, B.K.das, B. Adhikary, R.Verma, A.Shyam. Microwave emission from an AXIAL-Virtual Cathode Oscillator driven by a compact pulsed power source. February 2012. Conference: Journal of Physics: Conference Series 390 (2012) 012033 International Symposium on Vacuum Science  Technology and its Application for Accelerators. Volume: 390. DOI: 10.1088/1742-6596/390/1/012033.

\noindent [14]	P.T Chupikov, R.J. Faehl, I.N. Onishcenko, Y.V. Prokopenko, S.S. Pushkarev. Vircator efficiency enhancement assisted by plasma. IEEE Transactions on Plasma Science 34(1):14 – 17. 2006. DOI: 10.1109/TPS.2005.863590.

\noindent [15]	Hao Wu, Zhenghao He, Runkai Guo, Jun Ma, Huan Zhao, Xi Chen, Weimin Chen, Taisheng Xu. Characterization of the high current triggered vacuum switch. Vacuum.Volume 86, Issue 12, 20 July 2012, Pages 1911-1915. https://doi.org/10.1016/j.vacuum.2012.03.051.

\noindent [16]	Gang Lu; Minfu Liao; Yifan Sun; Ming Zhang; Qingxin Meng; Xiongying Duan. Research on Current-Rise Rate During Closing Process of Laser-Triggered Vacuum Switch.  IEEE Transactions on Plasma Science ( Volume: 50, Issue: 10, October 2022). DOI: 10.1109/TPS.2022.3204721.

\noindent [17]	Yingjie Fu, Mingtian Ye, Bingyang Feng, Luokang Dong, Shenyi Qin, Zhengzheng Liu, Mengbing He. Closing mechanism of electrically-triggered vacuum surface flashover switch. Vacuum. Volume 226, August 2024, 113273. https://doi.org/10.1016/j.vacuum.2024.113273.

\noindent [18]	Gang Lu a, Minfu Liao a, Ming Zhang a, Yifan Sun a, Jian Ou a, Xiongying Duan a, Xiaotao Han b. Calculation and analysis of plasma impedance during closing process of laser-triggered vacuum switch. Vacuum. Volume 205, November 2022, 111412. https://doi.org/10.1016/j.vacuum.2022.111412.

\noindent [19]	Ming Zhang a, Minfu Liao a, Gang Lu a, Hui Ma a, Qingxin Meng a, Xiongying Duan a, Xiaotao Han. Research on the trigger process of surface flashover triggered vacuum switch. Vacuum. Volume 211, May 2023, 111924. https://doi.org/10.1016/j.vacuum.2023.111924.

\noindent [20]	Zhanqing Chen, Xiongying Duan, Minfu Liao, Jiyan Zou. Closing performances of double-gap laser-triggered vacuum switch. October 2020.High Voltage 6(2). DOI: 10.1049/hve2.12026.

\noindent [21]	Ming Zhang, Minfu Liao, Liang Bu, Xiongying Duan. Research on Conduction Delay Time Characteristics of Double-Gap Surface Flashover Triggered Multistage Vacuum Switch. September 2024. Energies 17(18):4656. DOI: 10.3390/en17184656.

\noindent [22]	O.V. Manuilenko, I.N. Onishchenko, A.V. Pashchenko, I.A. Pashchenko, V.B. Yuferov. Current flow dynamics in plasma opening switch. // Problems of Atomic Science and Technology.2021,N4(134). P.6-10.

\noindent [23]	Magda, A. Pashchenko, et al. Modification of the Child-Langmuir-Boguslavsky law for the diode gap in the system with virtual cathode // Problems of Atomic Science and Technology. Series “Nuclear Physics Investigations”. 2012, N4, p. 133-137. 9.

\noindent [24].	A. Pashchenko, V. Ostroushko. Electron flow stability in the gas filled diode // Problems of Atomic Science and Technology. Series “Nuclear Physics Investigations”. 2017, N3, p. 72-76.

\noindent [25].	Mae Almansoori, Ernesto Neira, Sebastien Lallechere, Chaouki Kasmi, Felix Vega, Fahad Alyafei, "Assessing Vircators’ Reliability Through Uncertainty and Sensitivity Analyses Using a Surrogate Model", IEEE Access, vol.8, pp.205022-205033, 2020. DOI: 10.1109/ACCESS.2020.3037347.

\noindent [26].	Ernesto Neira, Felix Vega, Chaouki Kasmi, Fahad AlYafei, "Power Capabilities of Vircators: A Comparison between Simulations, Experiments, and Theory", 2020 IEEE 21st International Conference on Vacuum Electronics (IVEC), pp.313-314, 2020. DOI: 10.1109/IVEC45766.2020.9520518.

\noindent [27].	Patrizia Livreri. A Novel High-Efficiency S-Band Conical Axial Vircator.  IEEE Access PP(99):1-1. 2024. DOI: 10.1109/ACCESS.2024.3443108.

\noindent [28].	Giacomo Migliore , Antonino Muratore, Alessandro Busacca, Pasquale Cusumano, Salvatore Stivala. Novel Configuration for a C-Band Axial Vircator With High Output Power. IEEE transactions on electron devices, VOL. 69, NO. 8, AUGUST 2022. P. -4579-4585. DOI: 10.1109/TED.2022.3184917.

\noindent [29].	Schumer, J. W., Swanecamp, S. B., Ottinger, P. F., Commisso, R. J., Weber, B. V., Smithe, D. N. Ludeking, L. D. MHD-to-PIC transition for modeling of conduction and opening in a plasma opening switch. IEEE Trans. 2001 Plasma Sci. 29 (3), 479–493.CrossRefGoogle Scholar.

\noindent [30].	Shen Shou Max Chung, Shih Chung Tuan. C312: Impedance of Plasma Erosion Opening Switch in Particle-In-Cell Simulation. The 10th Euro-Asian Pulsed Power Conference (EAPPC 2024), The 25th International Conference on High-Power Particle Beams (BEAMS 2024) and The 20th International Symposium on Electromagnetic Launch Technology (EML 2024), At: Amsterdam, Netherland.

\noindent [31].	Mary Ann Sweeney, D.H. McDaniel, C.W. Mendel, C.D. McDougal. Plasma opening switch experiments on the Particle Beam AcceleratorII   Plasma Science, 1989. IEEE Conference Record - Abstracts., 1989 IEEE International Conference on. DOI: 10.1109/PLASMA.1989.165972.

\noindent [32].	Bruce Weber, R. J. Commisso, Phillip J. Goodrich, P. F. Ottinger. Investigation of Plasma Opening Switch Conduction and Opening Mechanisms. November 1991. IEEE Transactions on Plasma Science 19(5):757 – 766. DOI: 10.1109/27.108411.

\noindent [33].	D. P. Jackson, M. E. Savage, M. Gilmore, R. Sharpe. Overview of recent results from the triggered plasma opening switch experiment. January 2004. IEEE International Conference on Plas. DOI:10.1109/PLASMA.2004.1340116.

\noindent [34].	Limin Li, Guoxin Cheng, Le Zhang, Xiang Ji, Lei Chang et al. Role of the rise rate of beam current in the microwave radiation of vircator Role of the rise rate of beam current in the microwave radiation of vircator. April 2011. Journal of Applied Physics 109(7). DOI: 10.1063/1.3560774.

\noindent [35].	G. Liziakin,a) O. Belozerov, J. G. Leopold, Y. Hadas, and Ya. E. Krasik. Revisiting the Luce diode in the context of recent research on multi-vircators with dielectric reflectors. Physics of Plasma 31(3).(2024). DOI: 10.1063/5.0199092.

\noindent [36].	D. R. Welch, D. V. Rose, D. V. Novikov, M. E. Weller, A. A. Esaulov, G. H. Miley. Dynamics of the super pinch electron beam and fusion energy perspective. Physical review accelerators and beams 24, 120401 (2021).

\end{multicols}

\end{document}